\documentclass[
aps,pra,
reprint,
superscriptaddress,
amsmath,
amssymb, twocolumns]{revtex4-2}

\usepackage[utf8]{inputenc}
\usepackage[T1]{fontenc}
\usepackage{times}
\usepackage{microtype}
\usepackage{graphicx}
\usepackage{upgreek}
\usepackage{bm}
\usepackage{xcolor}
\usepackage[normalem]{ulem}
\usepackage{placeins}
\usepackage{amsmath}
\usepackage{amsthm}
\usepackage{dcolumn}
\usepackage{comment}

\usepackage{tcolorbox}
\usepackage{ragged2e}
\usepackage{caption}
\tcbset{colback=blue!7!white,colframe=gray!25!white,width=1\textwidth, 
  left=.8mm, right=.8mm, top=.8mm, bottom=.8mm}  
  
\usepackage{csquotes}
\usepackage[german,ngerman,english]{babel}
\usepackage{centernot}
\usepackage{mathtools}
\usepackage{ifthen}
\usepackage{tikz}
\usetikzlibrary{calc}
\usetikzlibrary{decorations.pathreplacing}
\usetikzlibrary{decorations}
\usetikzlibrary{positioning}
\usepackage{pgfplots}
\pgfplotsset{compat=1.18}
\usepackage{xxcolor}
\definecolor{structure}{rgb}{0.23,0.4,0.7}
\usepackage[normalem]{ulem}
\usepackage[colorlinks=true,linkcolor=teal,citecolor=teal,urlcolor=teal]{hyperref}

\newsavebox{\blocksavebox}

\definecolor{niceblue}{rgb}{0.33,0.5,0.8}

\newcommand{\refsub}[2]{\hyperref[#1]{\ref*{#1}#2}}

\newcommand{\norm}[2][]{
  \ifthenelse{\equal{#1}{}}
    {\left\| {#2} \right\|}
    {\ifthenelse{\equal{#1}{uinv}}
      {\left\vert\kern-0.25ex\left\vert\kern-0.25ex\left\vert {#2} \right\vert\kern-0.25ex\right\vert\kern-0.25ex\right\vert}
      {\left\| {#2} \right\|_{#1}}
    }
}

\newcommand{\taverage}[2][]{
  \ifthenelse{\equal{#1}{}}
  {\overline{#2}}
  {\overline{#2}^{#1}}
}

\newcommand{\tracedistance}[3][]{
  \ifthenelse{\equal{#2}{}}
  {\ifthenelse{\equal{#3}{}}
    {\mathcal{D}_{#1}}{}
  }{
    \ifthenelse{\equal{#1}{}}
    {\mathchoice{\operatorname{\mathcal{D}}\left(#2,#3\right)}{\operatorname{\mathcal{D}}(#2,#3)}{\operatorname{\mathcal{D}}(#2,#3)}{\operatorname{\mathcal{D}}(#2,#3)}}
    {\mathchoice{\operatorname{\mathcal{D}}_{#1}\left(#2,#3\right)}{\operatorname{\mathcal{D}}_{#1}(#2,#3)}{\operatorname{\mathcal{D}}_{#1}(#2,#3)}{\operatorname{\mathcal{D}}_{#1}(#2,#3)}}
  }
}
\newcommand{\fidelity}[3][]{
  \ifthenelse{\equal{#2}{}}
  {\ifthenelse{\equal{#3}{}}
    {\mathcal{F}_{#1}}{}
  }{
    \ifthenelse{\equal{#1}{}}
    {\mathchoice{\operatorname{\mathcal{F}}\left(#2,#3\right)}{\operatorname{\mathcal{F}}(#2,#3)}{\operatorname{\mathcal{F}}(#2,#3)}{\operatorname{\mathcal{F}}(#2,#3)}}
    {\mathchoice{\operatorname{\mathcal{F}}_{#1}\left(#2,#3\right)}{\operatorname{\mathcal{F}}_{#1}(#2,#3)}{\operatorname{\mathcal{F}}_{#1}(#2,#3)}{\operatorname{\mathcal{F}}_{#1}(#2,#3)}}
  }
}

\newcommand{\Sr}[3][]{
  \ifthenelse{\equal{#1}{}}
    {\operatorname{\mathnormal{S}}(#2\|#3)}
    {\operatorname{\mathnormal{S}}_{#1}(#2\|#3)}
}

\DeclareMathOperator{\1}{\mathbb{I}}



\definecolor{jens}{rgb}{0.1,0.5,0.1}
\definecolor{martin}{rgb}{0,0,1.0}

\newcommand{\jp}[1]{{\color{black} #1}}
\newcommand{\new}[1]{{\color{black} #1}}

\newcommand{\beq}[0]{\begin{equation}}
\newcommand{\eeq}[0]{\end{equation}}

\newcommand{\hide}[1]{}

\begin{document}

\title{Mind the gaps: The fraught road to quantum advantage}

\author{Jens Eisert}
\affiliation{Dahlem Center for Complex Quantum Systems, Freie Universit{\"a}t Berlin, 14195 Berlin, Germany}
\affiliation{Helmholtz-Zentrum Berlin f{\"u}r Materialien und Energie, 14109 Berlin, Germany}
\affiliation{Fraunhofer Heinrich Hertz Institute, 10587 Berlin, Germany}

\author{John Preskill}
\affiliation{Institute for Quantum Information and Matter, California Institute of Technology, CA 91125, USA}
\affiliation{AWS Center for Quantum Computing, Pasadena, CA 91125, USA}

\begin{abstract}
Quantum computing is advancing rapidly, yet substantial gaps separate today's noisy intermediate-scale quantum (NISQ) devices from tomorrow's fault-tolerant application-scale quantum (FASQ) machines. We identify four related hurdles along the road ahead: (i) from error mitigation to 
active error detection and correction, (ii) from rudimentary error correction to scalable fault tolerance, (iii) from early heuristics to mature, verifiable algorithms, and (iv) from exploratory simulators to credible advantage in quantum simulation. Targeting these transitions will accelerate progress toward broadly useful quantum computing.
\end{abstract}

\maketitle

\section{Introduction}

Quantum computers harness principles of quantum physics to solve some problems dramatically faster than conventional computers. 
The potential power of quantum computing was recognized in the 1980s~%
\cite{feynman1982simulating}, 
and interest in the topic skyrocketed in the 1990s when algorithms were discovered that, in principle, would allow quantum computers to break cryptosystems that are widely used to protect the privacy of digital communication~\cite{Shor-1994}. 

These developments stirred experimental physicists to pursue quantum computing hardware, but for several decades experiment lagged far behind theory. 
This is now changing. 
So-called \emph{noisy intermediate-scale quantum} (NISQ) computers now exist of sufficient size and accuracy 
that some tasks performed by the quantum machine are too complex to be emulated by the most powerful available \new{conventional supercomputers}~\cite{preskill2018quantum,morvan2024phase,PhysRevX.15.021052,abanin2025constructive}. 
NISQ technology is an impressive feat of engineering and opens exciting opportunities to explore quantum physics in previously inaccessible regimes. 
But quantum computations that are practically useful and economically 
viable have not yet been achieved, nor is it clear when that might 
happen. 

\begin{quote}
{\it Quantum hardware platforms are advancing steadily, but no one knows when quantum computers will run applications that broadly benefit society.} 
\end{quote}

The promise of quantum computing has attracted enthusiastic attention from companies, investors, and governments, as well as scientists, fueling the progress of NISQ technology. 
But NISQ computers are not error corrected, and the relatively high gate error rates of today's NISQ devices severely limit their computational power; therefore, 
\new{no immediate} 
practical uses of this technology have emerged to date. It has long been known how to overcome this limitation~\cite{shor1996fault}.
%
NISQ technology will be superseded by \emph{fault-tolerant application-scale quantum} (FASQ) computers that exploit quantum error-correcting codes to run a wide variety of useful applications. 
Quantum hardware has reached a sufficiently advanced stage that the voyage from NISQ to FASQ is gathering momentum in companies and research laboratories. We can envision fault-tolerant quantum computers with logical gate error rates that are many orders of magnitude better than the underlying physical gate error rates of hardware platforms. 
But this transition, still nascent, will unfold gradually on a timescale that remains highly uncertain, \new{primarily because building a large-scale quantum computer entails a colossal engineering challenge.}

\begin{quote}
{\it The path from NISQ to FASQ is likely to be arduous, expensive, and prolonged.}
\end{quote}

The divide between NISQ and FASQ is a particularly daunting gap that must be bridged to realize the full potential of quantum computing. 
In this perspective article, we highlight this and several other gaps that require the attention of the quantum community. 
Recognizing these gaps, and how to cross them, clarifies the challenges we face and the roadmap for gradual future progress.
%

It is helpful, for example, to divide the quest for FASQ into two stages. 
%
In the first stage (see Sec.~\ref{sec:mitigation}), we will cross the gap from today's error mitigation methods that enhance the power of NISQ computers to error-corrected devices that protect quantum information much longer than the inherent lifetime of the underlying physical qubits. 
The second stage (see Sec.~\ref{sec:fault-tolerant}) will build on these extended quantum memory times to realize universal fault-tolerant quantum computers that can solve hard problems. 
The first milestone will soon be within reach; the second will take longer. 
In fact, future quantum processors with broad utility may be based on different hardware modalities and/or architectural principles than those that are now being avidly pursued. 

A proposed application of quantum computing should satisfy three criteria. 
First, we desire an algorithm that runs efficiently on a quantum machine with quantifiable and reachable resource requirements. 
Second, we want a persuasive argument, possibly based on unproven but reasonable assumptions, indicating that any conceivable classical algorithm that achieves the same task has a much longer runtime than the quantum algorithm. 
And third, the quantum algorithm should provide a useful answer to a question that someone cares about. 
This answer might fuel scientific progress, or might provide economic value for companies, or might broadly benefit humanity. 
Whoever benefits, the answer should have intrinsic value, whether or not it was produced by a quantum machine. 
%
\begin{quote}
{\it We seek applications that are quantumly easy, classically hard, and practically useful.}
\end{quote}

The quantum community has proposed a few concrete ideas about where to find quantum utility, but substantial gaps in our knowledge must be addressed to make a compelling case that quantum computing will have significant practical value. 
One hope is that quantum machines will speed up the search for solutions to optimization problems and enhance the efficacy of machine learning (see Sec.~\ref{sec:heuristic}). 
But there is large gap between such aspirations and what has so far been rigorously established. 

Another proposal is that simulations using quantum computers will unveil static and dynamical properties of many-particle quantum systems in a regime beyond the reach of classical machines or previous quantum experiments. 
This aim may already be realized in the NISQ or early fault-tolerant eras (see Sec.~\ref{sec:simulation}). 
But for a while the results of such simulations will be most valued by scientists rather than for-profit industries. 
Eventually, quantum simulations may guide the discovery of new chemical compounds and exotic materials with significant societal value, but by some current estimates such applications have quantum resource requirements that might not be attainable soon. 
Our expectation is that scientific exploration enabled by near-term quantum computers will form the foundation for a variety of unforeseen applications in the longer term, just as happened in the case of conventional computing.

\begin{quote}
{\it 
Early applications of quantum computing will be primarily scientific. Broader economic impact will eventually follow.}
\end{quote}

Quantum utility will unfold gradually, fueled by advances in both technology and theory. 
Today's most capable NISQ machines can execute computations with fewer than $10^4$ two-qubit quantum operations. By some estimates, a broadly useful FASQ machine will need 
to be able to execute about $10^{12}$ such operations (a \textit{teraquop}) or more~\cite{lee2021even,gidney2021factor2}. As gate error rates 
improve and physical qubit counts increase, fault-tolerant quantum 
computing will pass 
through the \textit{megaquop} regime 
($\sim 10^6$ operations) 
and 
the \textit{gigaquop} 
regime ($\sim 10^9$ 
operations)~\cite{Megaquop}.
%
%
At each 
waypoint 
along 
the road from NISQ to FASQ, 
the quantum community will 
be challenged to explore the expanding application space enabled by the hardware progress. 

Several insightful recent articles 
\cite{zimboras2025myths,huang2025vast} 
provide valuable perspective on the quest for useful applications of quantum technology,  
and Ref.\ \cite{dalzell2023quantum} provides a useful overview of currently known quantum algorithms. Our article covers some of the same ground, with a particular emphasis on the 
four gaps along the road from NISQ that are highlighted above. 

We are optimistic about the quantum future, and our emphasis on the gaps faced by the quantum community is not intended to be discouraging. But only by adopting a clear-eyed assessment of the gaps that must be surmounted can we navigate the path to practical quantum utility that benefits the world. 

\newpage
\section{Quantum error mitigation and beyond}
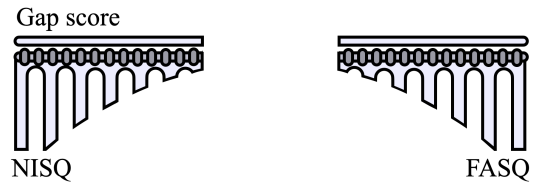
\label{sec:mitigation}


\begin{figure*}
\begin{tcolorbox}
\justifying{\noindent
\textbf{Box 1 | Leading quantum-computing platforms and their 
trade-offs}

\smallskip
The most advanced quantum processors today are based on trapped ions, superconducting circuits, and neutral atoms. All three platforms have achieved impressive control at the hundred-qubit scale, but differ markedly in gate speed, connectivity, and technological requirements such as fabrication, control, calibration, and system integration. These differences strongly influence error mitigation strategies, near-term applications, and fault-tolerant architectures.

\smallskip
\textbf{Trapped ions.}
In ion traps~\cite{IonTrapReview,IonQ,PhysRevX.15.021052}, each qubit is encoded in a single electrically charged atom, with state preparation, readout, and single-qubit gates implemented using laser pulses. Entangling gates exploit collective vibrational modes of the ions and typically require tens of microseconds. Key advantages are long coherence times, high-fidelity operations, and non-local connectivity enabled by ion movement. Drawbacks include slow gate speeds and further delays caused by the limited speed of ion transport. In large processors, multiple traps might be assembled into modular architectures connected either optically or by ion shuttling.

\smallskip
\textbf{Superconducting circuits.}
In superconducting processors~\cite{GoogleUnderThresholdCodes,FutureSuperconductingChips}, 
qubits are typically ``transmons,'' artificial atoms that are carefully fabricated and frequently calibrated, arranged in a two-dimensional grid with nearest-neighbor coupling. Microwave pulses enable fast single- and two-qubit gates, with entangling operations executed in tens of nanoseconds.
Coherence times may be improved through advances in materials 
and fabrication and/or by using alternative qubit designs (see Sec.~\ref{sec:fault-tolerant}). \new{The restriction to nearest-neighbor connectivity is a disadvantage compared to trapped-ion and neutral-atom processors.} Prospects for scaling to large processors might be enhanced by architectures that reduce the number of wires needed to control the device. 

\smallskip
\textbf{Neutral atoms.}
Neutral-atom platforms~\cite{LukinQEC,RevModPhys.82.2313,LukinQuantumGates} encode qubits in atoms held by optical tweezers or optical lattices.
Entangling gates in tweezer arrays are achieved on sub-microsecond timescales by driving pairs of atoms to highly excited Rydberg states with strong dipole interactions. Reconfigurable geometries enable flexible connectivity, 
\new{an important advantage,}
but slow atomic motion, measurement latency, 
and the need for continuous reloading pose challenges for reaching deep, fault-tolerant circuits.
Optical-lattice implementations using controlled atomic collisions provide an alternative approach~\cite{CollisionalQuantumGates}.

\smallskip
Across all three platforms, quantum circuits involving tens to hundreds of qubits have been demonstrated, with two-qubit gate error rates below 0.5\% and approaching 0.1\% in some cases \cite{FutureSuperconductingChips,PhysRevX.15.021052,ColdAtomsFidelities}.
Continuing improvements in hardware performance will extend the reach of error mitigation on NISQ processors, and will eventually reduce the overhead cost of fault-tolerant quantum computing.


}
\end{tcolorbox}
\end{figure*}

\jp{Today's quantum processors can execute quantum circuits with thousands of entangling operations acting on about 100 physical qubits.} Google Quantum AI has extracted a usable signal in quantum random circuit sampling tasks with 103 qubits and 40 layers of two-qubit gates using their Willow processor~\cite{GoogleUnderThresholdCodes}
(\new{compare}
also Refs.\ \cite{morvan2024phase,gao2025establishing}). 

IBM Quantum has reported successful execution of certain mirrored kicked Ising quantum circuits with up to 5000 gates~\cite{IBMnewsroom2024} using their Heron processor (see also
Ref.\ \cite{kim2023evidence}). These experiments helpfully benchmark the current status of superconducting quantum platforms. Given today's two-qubit gate error rates, circuits at this scale need to be sampled many times in succession to obtain an informative outcome, where the required number of repetitions scales exponentially with the circuit volume.

This exponential cost in 
sampling overhead is fundamental and unavoidable for noisy quantum circuits that are not protected by quantum error correction~\cite{ErrorMitigationObstructions,schuster2024polynomial,Nonunital,PRXQuantum.3.040329}.
Indeed, under depolarizing noise, as one applies additional layers of quantum gates without performing measurements and feedback, 
quantum states converge (in trace distance) to the maximally mixed quantum state \cite{PRXQuantum.3.040329}. When the qubit number $n$ is asymptotically large and the circuit depth scales like $\log n$, efficient classical Gibbs sampling algorithms accurately estimate output expectation values, thereby ruling out any quantum advantage \cite{FG20}. For non-unital noise such as amplitude damping, the noise itself can helpfully remove entropy arising from errors \cite{B01BenOr}, 
but nevertheless log-depth \emph{random} quantum circuits can be efficiently simulated classically on average~\cite{Nonunital,EPFLSimulation}. 

A variety of \textit{quantum error mitigation} (QEM) schemes can boost the reachable circuit volume significantly~\cite{RevModPhys.95.045005}. In QEM one samples from a family of quantum circuits and then applies classical postprocessing to the outcomes. For example, in \emph{zero-noise extrapolation} (ZNE) the noise strength in a circuit is varied artificially, and postprocessing extrapolates the results to the limit of zero noise~\cite{ZNE}. In \emph{probabilistic error cancellation} (PEC), the noise is characterized experimentally, and postprocessing inverts the noise process to recover an unbiased estimator for the circuit outcome~\cite{PhysRevA.104.052607}. 

Using \emph{subspace expansion methods}, errors are mitigated by post-processing a noisy quantum state within a carefully chosen low-dimensional subspace \cite{mcclean2020decoding}. Mitigation strategies can also reduce bias arising from measurement errors \cite{maciejewski2020mitigation}. QEM methods are essential for extracting valid results from today's NISQ processors \new{in the regime where the product of the gate count and the error per gate is of the order of unity}, but they become impractical for deep quantum circuits because the required sampling overhead scales unfavorably with circuit size \cite{ErrorMitigationObstructions,ErrorMitigationObstructionsOld}; 
%
in the worst case, the overhead is exponential in the product of the depth and width of the quantum circuit, 


\begin{quote}
{\it Quantum error mitigation boosts substantially the circuit volume that can be executed accurately.
However, due to a sampling overhead cost that scales exponentially with circuit volume, this method fails for very large circuits.} 
\end{quote}

Using suitable QEM schemes, circuits with 10,000 or more gates might soon be within reach, even without notable improvements in hardware quality~\cite{aharonov2025importance}\new{.} 
%
Circuits with, say, width 100 and depth 100 might, for example,  
\new{provide} scientifically valuable insights into the dynamics of many-particle 
quantum systems far from equilibrium (see Sec.~\ref{sec:simulation}).

\begin{quote}    
{\it Thanks to the substantial boost in circuit volume attainable via quantum error mitigation, 
NISQ machines with sufficiently low gate error rates might achieve marginally useful 
quantum advantage.}
\end{quote}


In contemplating the prospects for quantum advantage using NISQ devices, other factors 
besides circuit width and depth deserve attention.
\jp{In particular, gate speed, qubit connectivity, and measurement latency vary widely across today’s leading hardware platforms, shaping both the effectiveness of quantum error mitigation and the range of feasible applications (Box~1).} For example, ion traps and neutral atom devices have higher connectivity than today's superconducting processors, expanding 
the potential application space \new{and reducing the overhead cost of quantum error correction}. 
Importantly, superconducting circuits can sample more circuits because of their higher speed, enhancing the effectiveness of QEM.
Strictly speaking, rigorous lower bounds on the sampling overhead of QEM scale exponentially with the volume of the backward lightcone of the measured observable \cite{ErrorMitigationObstructions}, which can be much smaller than the total circuit volume in the case of low-depth geometrically local circuits.
We also note that QEM schemes 
\cite{PRXQuantum.4.010303,PhysRevX.11.041036,PhysRevX.11.031057}
that operate coherently on multiple quantum inputs have been proposed; these are not yet practical but may eventually prove to be valuable. 

Both \emph{quantum error correction} (QEC) and QEM have a significant overhead cost, but the nature of that overhead is quite different in the two cases. To protect a quantum circuit from noise using quantum error correction and fault-tolerant protocols, many additional physical qubits and physical gates are needed (see Sec.~\ref{sec:fault-tolerant}). Asymptotically, the overhead cost of QEC scales favorably --- polylogarithmically in the volume of the circuit we wish to execute; even so, the number of physical qubits required is beyond the capacity of current hardware. On the other hand, QEM, which does not require any extra physical qubits, is feasible today, but the sampling overhead is exponential in the circuit volume. Therefore, while QEC will surely be needed to execute very deep circuits, QEM is essential for seeking quantum advantage before the fault-tolerant era. In fact, even as QEC matures, QEM will continue to be valued for expanding the circuit size fault-tolerant platforms can reach \cite{LogicalZeroNoiseExtrapolation,PRXQuantum.3.010345}, 
at the cost of enlarging the sampling overhead; \new{such techniques can also mitigate circuit compilation errors}  
\cite{PRXQuantum.3.010345}.

\begin{quote}    
{\it Quantum error mitigation will continue to be useful in the FASQ era.}
\end{quote}

\section{From protected quantum memory to scalable fault-tolerant quantum computation}
\label{sec:fault-tolerant}



The discovery of efficient protocols for fault-tolerant quantum computing is a fundamental advance in our understanding of the physical universe \cite{Roads,PreskillReliable}.
%
Assuming that the currently accepted principles of quantum physics are correct, scientists and engineers of the future will be able to build extraordinarily complex quantum systems and control them with exquisite precision. 
%

This theoretical insight is nearly 30 years old. The first quantum error-correcting codes were discovered in 1995 \cite{shor1996fault,SteaneCode}, and schemes for using these codes to protect quantum computations against noise were formulated the following year \cite{shor1996fault}. 
The theory of fault-tolerant quantum computation establishes that, if the error probabilities for all physical operations are less than a constant value called the \emph{accuracy threshold}, and if the errors are only weakly correlated, then error rates for protected logical operations can be made arbitrarily small and therefore arbitrarily long quantum computations can be executed reliably \cite{FaultTolerance,PreskillReliable}. 
To reduce the logical error rate we need to increase the number of physical qubits and the number of physical gates per logical operation, but this overhead cost is sufficiently mild that we can feel confident that large-scale quantum computations will eventually be feasible. 

\begin{quote}
{\it An imperfect quantum computing device can accurately simulate an ideal computation with $L$ logical gates using $\mathcal{O}(L~\text{polylog}~L)$ physical gates, if the noise is sufficiently weak and not too strongly correlated.}
\end{quote}

\begin{figure*}
\begin{tcolorbox}
\justifying{\noindent
\textbf{Box 2 | Why fault-tolerant quantum computing is expensive}

\smallskip
\emph{Quantum error correction} (QEC) 
\cite{Roads,PreskillReliable}
protects fragile quantum information against noise by encoding logical qubits into larger blocks of physical qubits \new{or bosonic modes}. In principle, this protection allows arbitrarily long quantum computations to be executed reliably, provided physical error rates are below a threshold value and errors correlations are sufficiently weak.

\smallskip
A quantum error-correcting code is commonly denoted ${[[}n,k,d{]]}$, where $n$ is the number of physical qubits, $k$ is the number of protected logical qubits, and $d$ is the code distance.
A distance-$d$ code can correct up to $(d-1)/2$ errors.
High encoding rate ($k/n$) and large relative distance ($d/n$) are desirable, but practical performance also depends heavily on other factors that determine how effectively encoded information can be protected and processed.

\smallskip
Error correction proceeds by repeatedly extracting \emph{error syndromes} using quantum circuits, followed by classical \emph{syndrome decoding} to infer likely errors and apply appropriate recovery operations.
Because syndrome extraction itself is noisy, it must be designed to limit error propagation, and decoding must be fast enough to keep pace with the quantum processor \cite{GoogleUnderThresholdCodes,RigettiQEC,Roads}. Moreover, implementing universal logical gates typically requires mid-circuit measurements and classical feedforward,
further 
increasing architectural and control complexity.

\smallskip
The overhead cost of QEC scales favorably in theory, but can be daunting in practice.
As a concrete illustration, consider the surface code, a leading approach to early fault tolerance because it tolerates relatively strong physical noise and requires only geometrically local operations in two dimensions 
\cite{TopologicalQuantumMemory}.
Numerical simulations \cite{fowler2013surface,Myths} indicate that the logical error probability per operation satisfies
\begin{equation}\label{eq:Plogical}
P_{\text{logical}} \approx 0.1\left(\frac{p_\text{phys}}{p_\text{thresh}}\right)^{(d+1)/2},
\end{equation}
where $p_\text{phys}$ is the physical two-qubit gate error rate, $p_\text{thresh}\approx10^{-2}$ is the accuracy threshold, and $d$ is the code distance.
For the surface code, each logical qubit requires $n=d^2$ physical qubits.

\smallskip
To achieve, for example, a logical error rate $P_\text{logical}=10^{-11}$, which is suitable for running a highly parallel algorithm on $10^3$ logical qubits for $10^8$ time steps, one needs $d\approx19$ if $p_\text{phys}=10^{-3}$.
This corresponds to roughly $10^3$ physical qubits per logical qubit once ancillary qubits and logical-gate overhead are included, or about $10^6$ physical qubits in total—far beyond the scale of current devices.

\smallskip
This large separation between physical and logical resources poses a daunting challenge on the road from early error-corrected devices to scalable fault-tolerant quantum computers.}

\end{tcolorbox}
\end{figure*}

\jp{Useful fault-tolerant quantum computations might require billions of logical operations across 
thousands of logical qubits. To reach quantum circuits at this scale, we will need many more 
physical qubits, substantially better physical gate error rates, or both (Box~2).}
Furthermore, the qubit overhead is not the whole story; in addition, QEC inflates the time needed to run an algorithm. This may happen because many rounds of (noisy) syndrome measurements are needed to diagnose errors reliably, and because the architecture may hinder the parallelism of the logical circuit. 

Recent progress in both quantum hardware and QEC theory has ignited a vibrant interaction between the two. The current state of the art is exemplified by recent results from the Google Quantum AI team using a superconducting processor \cite{GoogleUnderThresholdCodes}. 
In a demonstration of a protected quantum memory hosting a single logical qubit, they performed millions of consecutive rounds of surface-code error syndrome measurement, each lasting about a microsecond. 
Significantly, they found that the logical error rate per measurement round improves by a factor $\Lambda\approx 2$ as the code distance increases from 3 to 5 and again from 5 to 7, suggesting that further improvements should be achievable as the code block continues to increase in size. 
As the hardware advances, we may anticipate larger values of $\Lambda$, larger codes achieving much lower error rates, and eventually not only protected quantum memory but also logical two-qubit gates with much better fidelity than physical gates. 

\begin{quote}
{\it For decades, experimental progress on QEC lagged far behind theory, but this is starting to change.}
\end{quote}

Meanwhile, alternative families of \emph{quantum low-density parity-check} (qLDPC) codes have been 
developed that have a much higher encoding rate (ratio of number of logical qubits $k$ to number of physical qubits $n$) than the surface code \cite{TillichZemor}. 
Among these are ``good'' codes such that $k/n$ and the relative distance $d/n$ remain bounded away from zero as $n$ approaches infinity \cite{Panteleev,LDPCReview,LeverrierLDPC}. 
Aside from these asymptotic results, relatively small codes with potential near-term relevance have been discovered recently. 
An intriguing example is an $\new{{[[n,k,d]]}=}{[[}144,12,12{]]}$ code \cite{IBMLowOverhead} \new{(see Box~2)} in contrast to the surface code with the same distance and length which encodes only a single logical qubit, this code protects 12 logical qubits, a significant improvement in encoding efficiency.

The catch is that measuring error syndromes for these higher-rate codes requires geometrically non-local physical operations. 
In superconducting circuits and other solid-state platforms, long-distance connectivity is hard to achieve but might be attainable via sophisticated fabrication techniques. 
High-rate codes are a better fit to atomic-qubit platforms like Rydberg arrays or 
ion traps, in which high connectivity can be realized by rearranging atomic positions. 
In addition, to improve the overhead cost of fault-tolerant quantum computing it is not enough to reduce the number of physical qubits that encode each logical qubit; we also need schemes for executing universal logical gates acting on the protected qubits with an acceptable cost. 
Here, too, steady theoretical progress is occurring, partly due to better ideas for tracking how errors propagate through fault-tolerant circuits \cite{xu2024fastparallelizablelogicalcomputation,Oratomic,QGPU}. 
%

Thanks in part to efficient error correction and logical gates enabled by qubit movement, impressive demonstrations of fault-tolerant circuits have been achieved recently using atomic platforms. 
These include circuits with 48 logical qubits on a 280-qubit Rydberg tweezer array system \cite{LukinQEC}, 
and circuits with 12 logical qubits on a 56-qubit ion trap device \cite{reichardt2024demonstration}. 
However, so far these demonstrations are limited to just a few rounds of error syndrome measurement, and 
postselection is invoked to achieve low 
logical error rates. 
That is, circuit runs are discarded when errors are detected, a scheme that will not scale to large 
circuits.
Another issue is that the Rydberg platforms, due to delays caused by \new{relatively} slow atomic movement and slow qubit measurements, have a logical cycle time that is orders of magnitude longer than superconducting processors. 
To run deep circuits with all-to-all coupling enabled by atomic rearrangement, much faster movement is 
desired, as well as faster readout and the ability to continuously load fresh atoms to replace those that are lost \cite{chiu2025continuous}.
\new{Rapid progress is underway to address all of these challenges. Recent estimates \cite{Pinnacle,Oratomic} based on plausible hardware assumptions indicate that high-rate qLDPC codes, enabled by nonlocal connectivity in atomic processors, could reduce the qubit overhead cost of fault-tolerant quantum computing by orders of magnitude compared to architectures with only nearest-neighbor connectivity.}

Fast real-time syndrome decoding 
\cite{GoogleUnderThresholdCodes,RigettiQEC,Roads}
is necessary in fault-tolerant quantum computing because we must frequently perform logical operations that are conditioned on prior measurements of logical blocks. 
If it takes too long to decode the measurement outcomes, that will slow down the logical clock speed \cite{RevModPhys.87.307}. 
As the size of the quantum circuit being executed increases, and code blocks grow accordingly to reach lower logical error rates, this decoding task becomes harder. 
Hence, we will need better decoding methods and faster classical processing as quantum computing scales.

\begin{quote}
{\it An error-corrected quantum machine will actually be a hybrid system requiring substantial classical processing power to decode error syndromes.} 
\end{quote}

We have focused here on superconducting circuits, ion traps, and Rydberg tweezer arrays because these are the quantum computing modalities that are now sufficiently advanced for pioneering explorations of quantum error correction. 
But with the fault-tolerant era just beginning to dawn, it remains far from clear which modality has the best prospects for reaching broadly useful quantum computers. 
It is telling that just a few years ago Rydberg atom platforms were not widely appreciated as a promising quantum technology, yet are now leading the way toward early fault-tolerant algorithms. 
It is vital to continue nurturing a variety of hardware approaches, including those like photonic and spin qubits that may lag behind for now but could leap forward in the next few years.

\begin{quote}
{\it We do not know yet which quantum computing modalities will be best suited for scaling to large systems that solve hard problems.}

\end{quote}

To cross the formidable gap from hundreds of physical qubits now to millions of physical qubits in future devices will require heroic advances in systems engineering, but curiosity-driven fundamental research will also help to show the way. 
Aside from investigating novel modalities that might have unforeseen advantages, we should explore qubit encodings that have potential to lower physical gate error rates substantially, thus reducing the qubit count needed to attain very low logical error rates. 

In the arena of superconducting circuits, many variations on this theme are already under investigation. 
Today's most advanced superconducting quantum computing devices are based on transmons; these are the simplest superconducting qubits and their quality is steadily improving. 
But since a logical qubit with very low error rate will be a very complicated object due to the hefty overhead cost of quantum error correction, it might pay off to make the physical qubits more complicated if the resulting gain in performance of the physical gates makes it easier to realize logical qubits \cite{Megaquop}. 

For example, a fluxionium qubit is more complicated than a transmon, because its large inductance is achieved with an array of many Josephson junctions, but the resulting large anharmonicity enadebles particularly low two-qubit error rates \cite{PRXQuantum.5.040342}.
%
\new{Bosonic
cat} qubits are realized by two-photon dissipation applied to a microwave resonator, which requires complicated hardware, but this approach strongly suppresses bit-flip error rates (while mildly increasing phase-flip error rates), potentially reducing the overhead cost of error correction \cite{putterman2025hardware}.
In a dual-rail encoding, a single qubit is encoded using a pair of resonators or transmons; the advantage is that the most common errors are directly detectable, simplifying the error correction task \cite{levine2024demonstrating}. 
\new{Using the Gottesman-Kitaev-Preskill error-correction scheme, encoded qudits in bosonic resonators have reached lifetimes exceeding the resonator's natural decay time \cite{GKPBreakEven}.}
A particularly ambitious approach to lowering physical gate error rates, and 
one of the first to be proposed 
\cite{TopologicalQuantumMemory},
is encoding an intrinsically robust qubit in a topological material 
\cite{aghaee2025distinct}.
%
A substantial leap forward in physical gate fidelity achieved using any of these or other strategies could redefine the roadmap to scalable fault tolerance.

Continuing progress in quantum error correction and fault tolerance will soon enable quantum technology to enter the megaquop regime in which circuits with a million or more two-qubit gates can be executed reliably \cite{Megaquop}. 
A megaquop machine will be able to perform some tasks that are beyond the reach of classical, NISQ, or analog quantum devices. 
Experience with these early fault-tolerant quantum computers will guide efforts to scale up toward more capable systems, and inform the quest for useful applications. 

\begin{quote}
{\it Early fault-tolerant ``megaquop machines,'' capable of reliably executing one million or more quantum operations, will be able to perform some tasks that are beyond the reach of classical, NISQ, or analog quantum devices.}
\end{quote}

There are ample opportunities for both hardware breakthroughs and theory advances that could accelerate progress toward scalable fault tolerance. 
More efficient encoding schemes, decoding algorithms, and fault-tolerant universal gates customized for the various hardware platforms will be helpful. 
Fresh ideas about system architecture may reduce the daunting requirements for local control of qubits in solid-state platforms like superconducting circuits. 
Deeper conceptual insights into the general principles underlying fault-tolerant quantum computation may suggest entirely new paradigms that reframe the pursuit of more powerful quantum technologies. 
Fundamental research has brought us to the brink of the fault-tolerant era, and will continue to guide our path to FASQ technology with broad applications.

\begin{quote}
{\it 
To build and operate FASQ machines with practical applications, substantial advances in both systems engineering and fundamental science will be needed.}
\end{quote}
 
\section{From near-term quantum heuristics to mature quantum algorithms}
\label{sec:heuristic}


A variety of potentially useful quantum algorithms are already known that can run on FASQ machines \cite{MontanaroOverview,dalzell2023quantum}, and we hope to discover many more. 
Comparatively little is known about algorithms for machines in the intermediate \new{regime} between NISQ and FASQ. 
How might we bridge this gap and strengthen the case for near-term quantum utility?
%

Arguably, quantum computing demonstrations that surpass classical simulation algorithms have already been achieved in \emph{random circuit sampling experiments}. \new{These include geometrically local circuits on superconducting processors involving more than 100 qubits and over 40 layers of two-qubit gates \cite{SupremacyReview,GoogleSupremacy,Boixo,
BiggestRandomSampling,morvan2024phase,gao2025establishing},
and nonlocal circuits on trapped-ion processors involving more than 50 qubits and over 10 layers of two-qubit gates \cite{PhysRevX.15.021052}; in addition, sampling in the classically hard regime has been claimed for post-selected logical circuits on neutral-atom processors \cite{Bluvstein}.}
\jp{While these results represent an important scientific milestone, their practical utility remains limited.}

\begin{quote}
{\it 
Quantum random circuit sampling experiments demonstrate that today's NISQ quantum processors can perform some tasks that are beyond the reach of today's classical supercomputers; this is a notable milestone, but not of great practical interest.}
\end{quote}

\jp{Efforts to realize genuine quantum utility in the NISQ era have been 
largely focused on \emph{variational quantum algorithms} \cite{Variational,bharti_2021_noisy}, with potential applications in combinatorial optimization \cite{QAOA} and machine learning \cite{biamonte2017quantum}. Despite intense interest and some noteworthy progress, these efforts have encountered some obstacles summarized in Box~3.}

\begin{figure*}
\begin{tcolorbox}
\justifying{\noindent
\textbf{Box 3 | Why near-term quantum advantage is hard to establish}

\smallskip
For certain quantum sampling problems, widely believed complexity-theoretic assumptions imply that the output distribution of a quantum computation is hard to sample from using any classical algorithm.
Although these hardness claims are asymptotic and rely on stringent notions of distributional closeness, present-day superconducting processors have reached regimes where direct classical simulation of such experiments is infeasible. Random circuit sampling is of little practical interest except for the purpose of benchmarking the quantum computer’s performance, but this achievement is an important scientific milestone, and attempts to validate the classical hardness have stimulated sizable improvements in classical methods for simulating quantum circuits~\cite{PhysRevLett.129.090502}.

\smallskip
Efforts to identify more directly useful applications in the NISQ era have focused largely on \emph{variational quantum algorithms} (VQAs), hybrid schemes in which a quantum processor prepares and measures parametrized quantum states and a classical computer iteratively updates the parameters to optimize a cost function~\cite{Gradients}.
Whether such algorithms can outperform the best classical heuristics on practical problems remains an open question, but serious obstacles have been identified.

\smallskip
Highly expressive parametrized circuits often exhibit \emph{barren plateaus}, where cost-function gradients become exponentially small, impeding training \cite{BarrenPlateaus,ReviewBarrenPlateaus}.
Reducing expressivity can alleviate this problem, but tends to make the resulting circuits easier to simulate classically~\cite{NoiselessSimulation,Nonunital,PhysRevLett.131.100803,DoesBarrenSimulability}.
In addition, the optimization landscapes of variational quantum algorithms often contain numerous local minima that are difficult to escape~\cite{AnschuetzTraps}.
Together, these effects suggest a fundamental tension between trainability and quantum advantage.

\smallskip
Despite these challenges, the potential value of NISQ variational algorithms remains unresolved.
\new{Potentially} promising strategies include \emph{warm starts}, in which classical heuristics inform the initial parameters or input states of the quantum algorithm~\cite{Warmstart1,StrategiesForRunningQAOA}, 
and hybrid workflows where classical computations guide how the quantum algorithm seeks nearby states that achieve a better value of the cost function~\cite{StrategiesForRunningQAOA}.

\smallskip
Another constructive approach is identifying ``proof pockets'' that establish advantage for relevant subtasks or for instructive special cases. 
For example, in the \emph{quantum approximate optimization algorithm} (QAOA)~\cite{QAOA}, optimal variational parameters have been rigorously shown to concentrate~\cite{QAOAConcentration}; 
hence, finding near-optimal parameters for one particular problem instance can 
advise a warm start in solving other instances.
In addition, single-round QAOA provably outperforms classical algorithms for problems with suitable symmetries \cite{AshleyQAOA}, and recent results identify graph families for which polynomial-time QAOA can find cuts through the graph with a cut fraction (ratio of the number of cut edges to the total number of edges) exceeding what is known to be achievable by any subexponential-time classical algorithm \cite{farhi2025lower}.
Further progress has been made by reducing or eliminating expensive gradient estimation in favor of more efficient dissipative optimization procedures \cite{NathanWiebe}. 
\new{There is also active research underway on improvements to and applications of variants of computational primitives such as quantum phase estimation that are intermediate between the NISQ and FASQ regimes \cite{PhysRevLett.129.030503}.}

\smallskip
While convincing end-to-end quantum advantage for near-term algorithms has not yet been demonstrated, accumulating rigorous results for carefully chosen settings may ultimately clarify where, and how, useful advantages can emerge.}
\end{tcolorbox}
\end{figure*}

\begin{quote}
{\it Quantum advantage in variational quantum algorithms has not been firmly established, and potential obstructions have been identified.
} 
\end{quote}
%
%
%

Beyond the NISQ regime, one may contemplate quantum advantage in optimization using FASQ machines. 
For NP-hard \emph{combinatorial optimization problems}, Grover's algorithm entails a quadratic speedup in exhaustive search for a satisfying assignment, and can likewise quadratically speed up heuristic classical search algorithms \cite{OptimizationReview}. 
However, this advantage kicks in only for very large instances, especially when we take into account the much slower clock speed of FASQ technology compared with conventional computing. 
The Grover speedup may prove to be valuable in the long term, but probably not until many decades from now \cite{babbush2021focus}.

We do not expect quantum computers to find exact solutions to worst-case instances of NP-hard problems efficiently, but they might be able to find better approximate solutions than classical computers, or to speed up the search for good approximate solutions. 
Variational quantum algorithms running on FASQ machines might realize that kind of advantage, but they face similar challenges to those confronting NISQ computers, such as barren plateaus and suboptimal local minima. 
It is known that directly simulating the \new{\emph{quantum approximate optimization algorithm} (QAOA)} is classically hard \cite{farhi2016quantum}, but also that log-depth QAOA does not have an asymptotic quantum advantage for a large class of sparse combinatorial optimization problems \cite{chen2023local} 
(which does not rule out a practically useful advantage at modest depth). 
As an existence proof, one can construct examples of asymptotic quantum advantage in finding approximate solutions to some optimization problems such as integer programming \cite{OptimizationAdvantages,Szegedy,BuhrmanAdvantage}. 
This is done by mapping problems solvable via Shor's algorithm to optimization tasks, and then invoking notions of computational learning theory \cite{OptimizationAdvantages} or a variant of the theory of probabilistically checkable proofs to show that finding an approximate solution is classically hard \cite{Szegedy}. 
Oracle-based subexponential quantum advantages in adiabatic optimization have also been identified \cite{gilyen2021sub,SubexponentialOptimization}.

New paradigms for seeking quantum advantage have emerged recently \cite{yamakawa2024verifiable}. 
A particularly intriguing fresh approach is \emph{decoded quantum interferometry} \cite{DecodedQuantumInterferometry} 
%
%
(DQI), in which the quantum Fourier transform maps a problem that seems classically hard to a decoding problem that is easy to solve classically. 
In particular, DQI provides an efficient quantum algorithm for \emph{optimal polynomial intersection} (OPI) which achieves a better approximation ratio than any currently known classical algorithm for a variant of polynomial interpolation. 
Recent work indicates that directly simulating DQI is classically hard \cite{marwaha2025complexity}, but also that DQI provides no quantum advantage in unstructured combinatorial optimization problems \cite{anschuetz2025decoded}. 
The potential quantum advantage achieved by DQI for structured problems like OPI is currently under investigation, as are the possible practical applications of such an advantage. 

\emph{Artificial intelligence} (AI) is transforming both science and technology, but faces obstacles such as the time and electrical power needed to train machine learning models. 
Hence it is natural to consider the potential power of combining quantum computing with \emph{machine learning} (ML) \cite{biamonte2017quantum,RevModPhys.91.045002}. 
Quantum ML is particularly well suited to address problems in quantum many-particle physics that are presumed to be classically hard (see Sec.~\ref{sec:simulation}). 
But what are the prospects for quantum advantage in tasks to which ML is currently applied?

Though compelling empirical evidence validates the scientific and economic value of AI as practiced today, we lack a rigorous understanding of what classical ML can and cannot do efficiently, which makes it all the more challenging to assess potential quantum advantages. 
The exponentially large Hilbert space of a many-qubit quantum system encourages speculation that mapping classical data to that large feature space can significantly enhance learning \cite{TemmeML}, but an important caveat is that the high cost of loading a large classical data set into the quantum device may offset the advantage of quantumly processing the data \cite{aaronson2015read}. 
Another proposal, that quantum advantage in ML derives from the ability to perform linear algebra tasks efficiently in an exponentially large space \cite{kerenidis2016quantum}, was ``dequantized'' when efficient classical sampling algorithms were formulated to perform the same task \cite{tang2019quantum}. 
Furthermore, the data encountered in practical machine learning tasks tends to be noisy 
with little discernible structure, 
while 
known performant quantum algorithms typically solve \new{highly structured problems} \cite{StructureAaronson}. 
The barren plateau problem cited above also raises concerns about the cost of training quantum neural networks. 

For some carefully chosen learning problems, quantum advantage follows from 
widely accepted classical hardness assumptions. 
Examples of learning tasks for which quantum advantage has been rigorously established include generative modeling \cite{PACLearning} (learning to sample accurately from a target distribution), density modeling 
(learning to return probability weights for sample inputs), binary classification \cite{TemmeML} (learning to sort data into two classes), and identification \cite{Molteni} (learning to assign unique labels to sample inputs). 
Such results are mildly encouraging, but these models are highly contrived and therefore do not yet provide 
much helpful guidance about where practically useful learning advantages should be sought. 
On a more positive note, there are indications that quantum algorithms can speed up the training of some classical neural networks for realistic classical data, for example by accelerating stochastic gradient descent \cite{JunyuPruned} or by invoking quantum linear system solvers to train generative models. 

\begin{quote}
{\it The potential for useful quantum advantage in end-to-end machine learning tasks remains largely unknown. 
Speedups have been established for some cherry-picked examples, but robust quantum advantages for large families of instances remain elusive.}
\end{quote}

The quantum linear systems algorithm \cite{PhysRevLett.103.150502} solves inversion of a sparse and well-conditioned matrix in time scaling logarithmically with the size of the matrix, where both the input and output are quantum states. 
Furthermore, the runtime scales polylogarithmically in the precision of the solution \cite{ChildsKothariSomma}. 
This algorithm, together with adaptive-order finite-difference methods and spectral methods, can be applied to solve linear partial differential equations 
\cite{ChildsPartialDifferential}. 
Such algorithms may have real-life applications in the FASQ regime, but various open issues still need to be addressed, such as encoding suitable boundary conditions, applying effective preconditioners, \new{extracting} usable information from measurements of the output state, \new{and understanding better the limitations of classical methods.} 
Extending these methods to nonlinear and stochastic differential equations would significantly expand their utility 
%
\cite{liu2021efficient}
 
\begin{quote}
{\it Various quantum algorithms are known that can achieve quantum advantage in the FASQ regime, but finding useful applications to real-life problems remains challenging.}
\end{quote}

\section{Toward credible quantum advantage in quantum simulation}
\label{sec:simulation}


Dirac may have been the first to articulate clearly that quantum mechanics provides a theory of everything that matters in everyday life \cite{dirac1929quantum}. 
It is the Schr\"{o}dinger equation that governs many electrons interacting electromagnetically with each other and with atomic nuclei. 
In principle, this equation explains all of chemistry and materials science, and everything built on those foundations. 
Sadly, Dirac lamented, this equation is ``much too complicated to be soluble'' for systems with more than a few electrons. 
It took over fifty years before Feynman \cite{feynman1982simulating} proposed that a machine that computes quantum properties of many strongly interacting particles should be a quantum machine rather than a conventional computer. 

Dirac's claim that many-electron problems are too hard to solve (classically) is in some respects misleading. 
Heuristic classical algorithms for this problem, such as density functional theory, 
are often quite successful in cases of practical interest. 
Furthermore, as both algorithms and hardware continually advance, classical methods are conquering more and more challenging problems. 
Quantum computing targets the (now) relatively small ``strongly correlated'' corner of chemistry and materials research, where classical methods falter. 
Because strongly correlated matter is poorly understood at present, it is difficult to foresee clearly how quantum simulations might drive future progress in fundamental science. 

\begin{quote}
{\it It is challenging to argue convincingly that ground-state problems for particular Hamiltonians of physical interest are both classically hard and quantumly easy.}
\end{quote}

Meanwhile, classical artificial intelligence is driving progress in many scientific fields. 
For now, applications of AI to quantum matter are hampered by a dearth of training data in the strongly correlated regime. 
We may anticipate that simulations by future quantum machines will provide abundant training data, enhancing the power of classical AI to make useful predictions about strongly correlated matter that has not yet been observed in the laboratory \cite{huang2021power,huang2022provably}.

\jp{Such considerations motivate a closer look at where quantum simulation offers the most credible prospects for quantum advantage, and where classical methods are likely to remain competitive. 
Box~4 emphasizes that, while classical heuristics are often capable of capturing equilibrium properties of physically relevant Hamiltonians, they are markedly less effective at predicting nonequilibrium dynamical behavior; hence dynamics offers a more plausible route to quantum advantage.}

\begin{figure*}
\begin{tcolorbox}
\justifying{\noindent
\textbf{Box 4 | Where quantum simulation may achieve credible advantage}

\smallskip
Quantum simulations use controllable quantum systems to study the properties of quantum matter \cite{CiracZollerSimulation,BlochSimulation,BlattSimulator,ProgrammableSimulationIons}.
Simulations may target either static properties of equilibrium states or dynamical behavior far from equilibrium, and may be carried out using either digital gate-based quantum computers (\emph{digital quantum simulation}) or analog platforms with programmable Hamiltonians (\emph{analog quantum simulation}).
Both types of problems, and both types of quantum platforms, have significant scientific value.

\smallskip
A widely studied task is computing ground-state properties of local Hamiltonians.
Although exact classical solutions are feasible only for small systems, a variety of powerful heuristic classical methods—such as density functional theory and tensor-network approaches—often perform well in practice for physically relevant problems, particularly in low dimensions.
Quantum algorithms for ground-state preparation exist \cite{PhysRevA.109.042425,EfficientGSPrep}, but their success typically depends on the availability of suitable initial states or favorable energy landscapes. As a result, it has proven difficult to argue convincingly that ground-state problems of direct physical interest are both classically hard and quantumly easy \cite{lee2023evaluating}.

\smallskip
By contrast, simulating the \emph{dynamics} of quantum systems far from equilibrium may be a more promising route to quantum advantage.
Classical methods for dynamical simulation are generally less effective than those for equilibrium properties, and rapidly growing entanglement can render classical descriptions inefficient.
Quantum platforms can naturally realize such dynamics, whether through digital time evolution or the intrinsic evolution of analog simulators \cite{Trotzky,BlochSimulation,Dissipation}.

\smallskip
A typical quantum simulation involves preparing an initial state, evolving it under a target Hamiltonian, and measuring observables of interest, with many repetitions required to obtain accurate expectation values.
While digital simulation algorithms can in principle achieve favorable asymptotic scaling with system size and simulation time \cite{haah2021quantum} 
\new{and methods such as
quantum signal processing 
and 
qubitization}  \cite{Qubitization} have more favorable asymptotic costs},
their practical implementation may demand deep circuits and substantial overhead.
Analog simulators, which avoid much of this overhead, can access larger system sizes but offer less flexibility and weaker control over errors.

\smallskip
From a complexity-theoretic perspective, any problem efficiently solvable by a quantum computer can be mapped to the simulation of a local Hamiltonian, implying the existence of carefully constructed models for which quantum simulation is classically hard \cite{Vollbrecht}.
However, this observation does not by itself establish classical hardness for families of Hamiltonians arising naturally in physics.

\smallskip
Quantum simulation—especially of nonequilibrium dynamics—offers a compelling path toward scientifically meaningful quantum advantage, but demonstrating such advantage for physically relevant systems remains an open challenge.
\end{tcolorbox}
\end{figure*}


\jp{As noted in Box~4,} and as for quantum algorithms more broadly, we may anticipate an ongoing competition between quantum teams who run simulations on quantum hardware, and classical teams who try to match or surpass the results using conventional computers (often based on tensor-network, neural-network, or Pauli-path methods), a line of thought dating back to 2011 \cite{Trotzky}. 
Indeed, that competition is increasingly under way. For example, the results of a 
\new{stroboscopic}
quantum simulation of a two-dimensional 
kicked Ising model using 127 qubits on IBM hardware \cite{EvidenceUtility}, initially claimed to be a demonstration of quantum utility, were quickly matched by classical calculations using both tensor-network methods \cite{PRXQuantum.5.010308} and sparse Pauli dynamics \cite{beguvsic2023fast}. 
Results of an impressive simulation of rapidly quenched spin-glass dynamics using D-Wave hardware \cite{DwaveAdvantage}, also reported as beyond the reach of classical methods, have been partially reproduced using classical tensor-network simulations \cite{StoudenmireDWave}. 
This healthy back-and-forth between quantum and classical simulation teams is helpful and stimulating for both sides, but making a fair comparison involves some subtleties. 
Classical simulations are often limited to small system size, and quantum simulations have limited accuracy. 
Furthermore, more information can be extracted from the output of a classical simulation than from the output state of a quantum simulation, even when the quantum simulation is repeated many times. 

Although quantum simulations in the NISQ era may not have very high accuracy, they might still be qualitatively informative. A recent example illustrating the state of the art is a computation of \emph{out-of-time-order correlators} (OTOCs) using 103 qubits on a Google superconducting processor in a regime that is arguably hard to simulate classically \cite{abanin2025constructive}. 
To acquire quantitatively useful data, fault-tolerant quantum computers will most likely be needed, but useful quantum advantage might be attained in simulations of strongly-coupled lattice systems in two dimensions using relatively near-term digital machines in the megaquop to gigaquop regime. 
Thoughtful co-design of fault-tolerant architectures and simulation algorithms will help to ensure that useful simulations are reached sooner.

In contrast to a circuit-based digital universal quantum computer, an analog quantum simulator is a quantum system with many degrees of freedom that can be tuned to resemble a model system we wish to study and understand. Some of the same experimental platforms, for example Rydberg tweezer arrays \cite{BrowaeysRydberg,ProbingTopological,RevModPhys.82.2313}, other ultracold-atom platforms \cite{BlochSimulation,Trotzky}, superconducting circuits \cite{Dimitris}, and trapped ions \cite{ProgrammableSimulationIons,BlattSimulator}, can be used for both purposes. 
While interesting digital quantum simulations of many-particle quantum systems are just starting to become possible now, analog simulations of such systems have been studied productively for over two decades \cite{BlochSimulation,BlattSimulator,ProgrammableSimulationIons}. 

The fundamental elements of an analog simulator can include harmonic resonators or fermionic atoms, evading the overhead cost of simulating those elements using qubits. 
\new{(It is also noteworthy that the overhead cost of digital quantum simulation might be reduced using processors fashioned from bosonic or fermionic degrees of freedom.)}
In a particularly powerful analog approach, ultracold fermions may be trapped in optical lattices with single-site addressability \cite{DopedFermiHubbard}. 
%
These systems illuminate properties of strongly correlated systems that are difficult to access in conventional solid-state experiments, and for much larger system size than can be reached with today's digital quantum platforms. 
For example, phase diagrams of Hubbard-like models can be explored, and emergent hydrodynamic transport of spin and charge can be studied. 
Recent evidence suggests that fermionic models are intrinsically harder to simulate than spin models \cite{AnschuetzFermions}. 

Despite these undeniable advantages, analog platforms also have some drawbacks. 
Strongly correlated physics of interest may emerge only at very low temperatures that are hard to reach experimentally, and due to imperfect control the actual Hamiltonian in the lab may differ significantly from that of the target system. 
This issue might be partially mitigated by validating the Hamiltonian using Hamiltonian-learning algorithms \cite{PhysRevLett.112.190501,HamiltonianLearning}. 
We also note that the accuracy of an analog simulator may be considerably better than worst-case error bounds would indicate, because errors are likely to partially cancel out \cite{trivedi2024quantum}.
The digital approach affords greater flexibility in the choice of initial state and Hamiltonian, and admits quantum error correction for more robust control. 
As happened in conventional computing, analog quantum simulators may eventually become obsolete. 
However, given the hefty overhead cost of quantum fault tolerance, analog platforms will maintain notable discovery potential well into the future, especially in studies of quantum dynamics far from equilibrium. 

\begin{quote}
{\it Analog quantum simulators may be surpassed by digital quantum simulators eventually, but in the near term they are powerful tools for scientific exploration, well suited for studies of quantum dynamics.}
\end{quote}

Not only do we want quantum simulators to solve classically hard problems, we also want them to provide useful knowledge beyond what would otherwise be accessible. 
According to a recent estimate from a High Performance Computing facility operated by the U.S.\ Department of Energy \cite{
camps2025quantum}, over 30\% of compute cycles are spent on density functional theory computations for chemistry and materials science, over 18\% on lattice quantum chromodynamics, and about 2.6\% on chemistry algorithms other than DFT. 
These numbers provide a rough indication of the level of activity relating to quantum problems in the computational physical sciences, and do not include additional topics like molecular dynamics and fusion energy which also account for a large fraction of the computational workload and where more precise treatment of quantum effects might yield improved results. 

Classical computing will continue to be widely and productively used in this problem space, but we are optimistic that quantum simulation will provide surprising insights beyond the scope of classical methods. 
We anticipate discovering new highly entangled phases of matter, both in equilibrium and far from equilibrium, and exploring properties of strongly coupled quantum systems that are beyond the reach of other computational methods. 
We expect substantial impact first in condensed matter physics, later in chemistry, and eventually in high energy physics and even quantum gravity. 
While we are confident about the potential of quantum simulators as tools for scientific discovery, it is more difficult to predict their usefulness for industrial applications. 

\begin{quote}
{\it Quantum simulations on digital and analog quantum platforms 
will be scientifically informative, but their economic value is less clear.}
\end{quote}

\section{Outlook}

The rapid recent advance of quantum science and technology has drawn attention not just from scientists but also from commercial ventures and government policy makers. 
Quantum computers with broadly useful applications are eagerly anticipated but are challenging to realize. 
In this article we have emphasized that to navigate the road from today's NISQ era to tomorrow's FASQ machines the quantum community must ``mind the gaps'' that block the way. 
We have highlighted in particular the daunting quest for scalable fault-tolerant quantum computers and the crucial hunt for practical quantum advantages over conventional information processing. We stress, though, that the portfolio of useful applications is likely to expand gradually as quantum technology advances through the megaquop, gigaquop, and teraquop regimes. 

How will quantum technology advance science and benefit society? 
Efforts to envision the long-term impact of quantum technology on the world are hampered by our limited imaginations --- the history of classical computing vividly illustrates that predicting the future course of information technology is beyond our grasp. 
In a letter to Lewis Strauss in 1945, John von Neumann reflected on the potential value of high-speed electronic computers. After delineating some prospective applications, he wisely opined \cite{vonNeumann-1945-to-Strauss}:

\begin{quote}
{\it [T]he projected device\,[\dots] is so radically new that many of its uses will become clear only after it has been put into operation\,[\dots]. [T]hese uses which are not, or not easily, predictable now, are likely to be the most important ones. Indeed they are by definition those which we do not recognize at present because they are farthest removed from what is now 
 feasible, and they will therefore constitute the most surprising and farthest-going extension of our present sphere of action \dots}
\end{quote}

Just as von Neumann could not envision the digital revolution that would eventually unfold, we cannot hope to foresee today the most impactful future applications of quantum computers. 
Arguably our task is all the more hopeless because quantum information processing entails an even larger step beyond past experience. 
Despite our best efforts to predict the important applications, tomorrow's quantum computers are sure to delight and benefit us in ways we cannot currently anticipate. Before that happens, we have a lot of work to do.

\section*{Acknowledgments}

We are grateful to Steve 
T.~Flammia for helpful and insightful comments about this article, and to many other colleagues for years of stimulating discussions about the quest for quantum advantage. We also thank 
Ieva Cepaite,
Edward Farhi,
Mark Goh,
Sam Gutmann,
Peter Love,
Shaghayegh Moradirad,
and Zoltan Zimboras
for valuable comments.
The work of J.~E. has been supported by the BMFTR 
(QSolid, DAQC, MUNIQC-Atoms, QuSol, PasQuops, \new{Hybrid++}), the Munich Quantum Valley (K-4 and K-8), the Quantum Flagship (PasQuans2, Millenion), 
QuantERA (HQCC), the 
Clusters of Excellence MATH+ and ML4Q, the DFG (CRC 183, \new{SPP 2514}), the Einstein Foundation (Einstein Research Unit on Quantum Devices), Berlin Quantum, and the ERC (DebuQC). 
The work of J.~P. has been supported in part by the U.S. Department of Energy, Office of Science, National Quantum Information Science Research Centers, Quantum Systems Accelerator, by the U.S. Department of Energy, Office of Science, Accelerated Research in Quantum Computing, Quantum Utility through Advanced Computational Quantum Algorithms (QUACQ) and Fundamental Algorithmic Research toward Quantum Utility (FAR-Qu), and by the National Science Foundation (PHY-2317110). The Institute for Quantum Information and Matter is an NSF Physics Frontiers Center.\smallskip


%

\end{document}